\documentclass[twocolumn,showpacs,preprintnumbers,amsmath,amssymb]{revtex4}

\usepackage{graphicx}
\usepackage{dcolumn}
\usepackage{bm}

\begin{document}


\title{A Froggatt-Nielsen flavor model for  
neutrino physics}

\author{Micheal S. Berger}
 \email{berger@indiana.edu}
\author{Maria Dawid}%
 \email{mdawid@indiana.edu}
\affiliation{%
Physics Department, Indiana University, Bloomington, IN 47405, USA}
%


\begin{abstract}
Superheavy neutrinos can, via the seesaw model, provide a mechanism for lepton number violation. If they are combined with flavor violation as characterized by the 
Froggatt-Nielsen mechanism, then the phenomenology for the neutrinos in oscillation experiments, neutrinoless double beta decay, and other experiments can be described 
by a relatively few number of parameters. We describe the low-energy neutrino mass matrix and show that the results are consistent with currently available data.
\end{abstract}

\maketitle

\def\half{{\textstyle{1\over 2}}}

\section{Introduction}

An amusing property of the neutrino masses and mixing is that two mixing angles are 
large. An early expectation was that instead the mixing between flavors would mirror the small 
mixing observed for quarks and charge leptons. However, neutrino masses may not be Dirac masses
like their quark and charged lepton counterparts. They can obtain their 
small masses by virtue of a seesaw mechanism giving light neutrinos with Majorana masses, 
so perhaps there can be different opportunities 
for generating large mixing angles.

One large mixing angle accounts for the solar neutrino oscillations, wheras the other is involved in
the atmospheric neutrino data. Analysis of the flavor symmetry in the charged lepton and quark sectors 
of the Standard Model tries to account for the hierarchy of masses and the smallness of the mixing angles
observed. For example the Froggatt-Nielsen mechanism introduces a horizontal symmetry and a small
parameter, which characterizes the breaking of the symmetry, can be used to accommodate the pattern
of masses and mixings observed there. For this reason the Froggatt-Nielsen mechanism is 
usually considered as an avenue in flavor models for explaining small mixing angles and large mass
hierarchies, a pattern not observed in the neutrino sector. However,
for neutrino physics a combination of the seesaw mechanism 
with the Froggatt-Nielsen mechanism  gives rise to new possibilities for accounting for the 
neutrino data. 

\section{Definition of the Model}

In this paper we extend an earlier model\cite{Berger:2006zw} which is based on  a 
symmetry 
\begin{eqnarray}
G_{SM}\times U(1)_H\times U(1)_L\;,
\label{symmetry}
\end{eqnarray}
where $G_{SM}$ is the Standard Model gauge group, $U(1)_H$ is the usual
horizontal (or flavor) symmetry of Froggatt-Nielsen\cite{Froggatt:1978nt}
models, and $U(1)_L$ is 
lepton number. Scalar fields will be used to break the two $U(1)$ symmetries. The familiar
Froggatt-Nielsen scalar field $S_H$ breaks the horizontal symmetry.
Some other scalar fields which break the horizontal $U(1)_H$ will also carry lepton number.
Neutrino anarchy\cite{Hall:1999sn}
will be used to make the second and third generations of light neutrinos
with large mixing. This approach was used in Ref.~\cite{Berger:2006zw} which made the
then popular but now obsolete prediction of a vanishing value for the $U_{e3}$ matrix element
(i.e. a vanishing third mixing angle) of the light neutrino mixing matrix. This model prediction
occured because only 
two generations of superheavy neutrinos were needed to 
accommodate the neutrino data existing at that time. Now that a nonvanishing $U_{e3}$ is 
required by the neutrino experimental data, the model can be extended to include 
three (or more) generations of heavy neutrino to accommodate this measurement.

In Froggatt-Nielsen models there is a small parameter which gives rise to hierarchies in the
charged lepton and quark sectors.  In a fully anarchical three generational model the horizontal 
charges would be assigned to be the same for each of the three generations. Then the challenge
 is to understand small quantities in neutrino physics, like the ratio of $\Delta m^2$ observed in 
solar and atmospheric
neutrino oscillations as well as the smallness of the third mixing angle.
In the model described in this paper, on the other hand, the horizontal symmetry is again
broken using a small Froggatt-Nielsen parameter, but this small parameter is also used to 
suppress the small quantities mentioned above.
An agreement with the experimental neutrino 
oscillation data\cite{Tanabashi:2018oca} can be obtained in a very economical way.

A scalar field $S_{HL}$ which 
carries charge $-1$ under $U(1)_H$ (which generates the mass and mixing angle
hierarchies in the charged lepton and quark sectors of the Standard Model when the Froggatt-Nielsen
scalar field $S_H$ receives a vev). However,
the new field $S_{HL}$ also violates lepton number upon spontaneous symmetry
breaking of $U(1)_L$. All light neutrino Majorana masses must reflect this symmetry breaking.
In Ref.~\cite{Berger:2006zw} this was called ``combining'' the lepton 
number violation with the flavor symmetry violation, and the resulting model
exhibits a meshing of the seesaw and Frogatt-Nielsen mechanisms. It has been noted
before that there are interesting and more subtle possibilities for the Froggatt-Nielsen
mechanism in the neutrino sector which may involve a seesaw\cite{Nir:2004my,Nir:2004pw}.

A mechanism to obtain the large observables in neutrino physics is an $L_e-L_\mu -L_\tau$ symmetry\cite{Petcov:1982ya,Goh:2002nk}, which can be implemented in a seesaw model and 
be perturbatively broken by the Froggatt-Nielsen mechanism. Before including 
these small effects, the light neutrino mass matrix in the flavor basis has the form
\begin{eqnarray}
m^{}_{\ell 1}&=&
m\left (\begin{array}{ccc}
0 & \sin \theta & \cos \theta \\
\sin \theta & 0 & 0 \\
\cos \theta & 0 & 0 
\end{array}\right )\;.
\label{unperturb}
\end{eqnarray} 
This matrix has mass eigenvalues $0$, $-m$, and $+m$, so it is of the 
inverted hierarchy type.  It is diagonalized by a unitary matrix $U_\nu$
which contains a maximal mixing angle which is 
associated with solar neutrino oscillations (its origin is of the pseudo-Dirac type), 
and a large mixing angle ($\theta$)
which is responsible for atmospheric neutrino oscillations. i.e.
\begin{eqnarray}
\sin ^2 2\theta _A&=&\sin^2 2\theta\;, \nonumber \\
\sin ^2 2\theta_\odot &=& 1\;.
\label{unperturb_angles}
\end{eqnarray}
The second and third generations are assigned the same charges under the 
symmetries (neutrino anarchy). The first 
generation is assigned a different $U(1)_H$ charge, and the breaking of this
symmetry via the  
Froggatt-Nielsen mechanism, with interactions consistent with Eq.~(\ref{symmetry}),
are needed to fill in the zeros in this matrix. Therefore the spontaneous breaking of the 
$U(1)_H$ symmetry determines the size and structure of the perturbations.

The Pontecorvo-Maki-Nakagawa-Sakata (PMNS) matrix receives contributions from the nonalignment
of the flavors and masses in the charged lepton sector as well as the neutrino mass matrix.
The matrix is defined as
\begin{eqnarray}
U_{\rm PMNS}=U_L^\dagger U_\nu\;,
\end{eqnarray}
where $U_L$ is the matrix that diagonalizes the charged lepton sector. Since their exists 
a large hierarchy in the charged lepton ($e$, $\mu$, $\tau$) masses, Froggatt-Nielsen 
models should give small mixing angles in the charged lepton sector, and the matrix $U_L$ is 
approximately the identity.
Therefore the mixing angles arising from $U_\nu$ 
to provide the dominant contributions to the mixing angles observed in the 
experiments in $U_{\rm PMNS}$, and we henceforth restrict our attention to 
the diagonlization of this matrix.

In Ref.~\cite{Berger:2006zw}
we introduced a full three-generation model but with only two generations
of superheavy neutrinos. In this paper we expand the set of superheavy
neutrinos to include a full set of three generations. If this is done assuming
the second and third generations of heavy neutrinos have the same horizontal
charges (anarchy), then the model naturally accounts for all the available neutrino data: 
large mixing angles for solar and atmospheric neutrino oscillations as well
as a nonzero value
for $U_{e3}$ and the possibility of CP violation in the lepton sector of the 
Standard model. The model has an inverted mass hierarchy imposed by the 
$L_e-L_\mu -L_\tau$ symmetry, exactly one massless 
neutrino, and the possibility of an observable signal in neutrinoless double beta decay 
experiments. 

The scales in the model are: 

1) the electroweak breaking scale determined by 
the vevs of Higgs doublets, $\langle \phi_{u,d}\rangle$;

2) $M_{HL}\equiv \langle S_{HL}\rangle \sim \langle \bar{S}_{HL}\rangle$, the 
lepton number breaking scale;

3) $M_H\equiv \langle S_H\rangle \sim \langle S_H^\prime\rangle$, 
the horizontal symmetry breaking scale;

4) $M_F$, the mass scale of Froggatt-Nielsen vector-like quarks and leptons.

The Froggatt-Nielsen parameter 
\begin{eqnarray}
\lambda_H\equiv {{\langle S_H\rangle}\over M_F}={M_H\over M_F}\;,
\end{eqnarray} 
can be used to generate the mass and mixing angle hierarchies in the quark
and charged lepton sectors of the Standard Model.
It is small and is often associated with the value of Cabibbo angle ($\sim 0.2$)
since it is used to generate the mass and mixing angle hierarchies in the quark
and charged lepton sectors of the Standard Model (as mentioned above, to achieve
the hierarchy in masses and mixing needed to account for the data, usually 
large powers of the parameter is needed).
The lepton number breaking introduces another parameter 
\begin{eqnarray}
\lambda _L^2\equiv {{\langle S_H \rangle ^2}\over {\langle S_{HL}\rangle
\langle \bar{S}_{HL}\rangle}}={M_H^2\over M_{HL}^2}\;.
\end{eqnarray} 
This parameter, being the ratio of two vevs breaking the horizontal symmetry, may
or may not be small. In the case where $\lambda _L$ is small, the neutrinos
may display a mass hierarchy which is significantly different from that suggested by
their Froggatt-Nielsen $U(1)_H$ alone. Alternatively, as in the model presented in this 
paper, there can be additional symmetry considerations that constrain the model 
predictions. Our results will apply whether $\lambda _L \sim 1$, $\lambda_L \sim \lambda _H$, 
or even smaller.

The fields of the model and their associated $U(1)_H$
charges
\begin{eqnarray}
L_0, L_{+1}^{(k)}, N_{+1}^{(i)}, \bar{N}_{-1}^{(i)}, N_0^{(j)}, \bar{N}_0^{(j)} \;,
\label{def}
\end{eqnarray}
where $k=1,2$ to account for the three generations of lepton doublets $L$
 in the Standard Model. The fields $N$ are the Standard Model singlets. The fields
$L$ and $\bar{N}$ give rise to the usual Dirac mass terms. Vector-like 
couplings can arise between $N$ and $\bar{N}$ while lepton violating couplings
arise between two $N$ fields or two $\bar{N}$ fields. The indices $i,j$ indicate 
the possibility of multiple pairs of each horizontal charge. Any number of additional
pairs of superheavy neutrinos of each type will produce a model with the 
properties of the minimal viable model, $i=1$ and $j=1,2$. In Ref.~\cite{Berger:2006zw}
only one pair of each type ($i=1$ and $j=1$) was considered which produces the most predictive
and economical description
of the neutrino data (at the time). It turns out introducing a second pair 
$ N_0^{(2)}, \bar{N}_0^{(2)}$ is sufficient to produce a nonzero prediction for 
$U_{e3}$ entry in the PMNS matrix. Furthermore, this additional pair of 
leptons makes the field content of the model consistent over the three
generations.

The fermion fields have lepton number assignments as given in 
Table~\ref{lepton}. The $U(1)_L$ symmetry  
ensures lepton number $L=L_e+L_\mu+L_\tau$ conservation before 
symmetry breaking from vevs of the scalar fields. With $k=1,2$ we have the 
three Standard Model generations $L_{+1},L_0^{(1)},L_0^{(2)}$. In particular the 
charges of the heavy neutrinos $N_{+1}$ and $N_0^{(k)}$ have the same charges
as the light fields, $L_{+1}$ and $L_0^{(k)}$ respectively. The $\bar{N}$ fields have 
the opposite charges of their $N$ partners in all cases. For this reason, it is natural to 
suppose that $i=1$ and $j=1,2$ so that the heavy neutrinos fall into the same
set of generations. We present the phenomenological results for this case below,
but the more generic equations can be obtained by simple and obvious generalizations.

The scalar fields are assigned charges according to Table~\ref{scalar}.
The $L_e-L_\mu-L_\tau$ symmetry, respected by the $S_{HL}$ and $\bar{S}_{HL}$
fields, can be understood from the 
charges assigned to the fields in Eq.~(\ref{def}). These fields acquire vevs that violated
overall lepton number $L$, but do not violate $L_e-L_\mu-L_\tau$. The 
fields $S_H$ and $S_H^\prime$ also respect the $U(1)_L$ symmetry, but their interactions
violate $L_e-L_\mu-L_\tau$ conservation. The small Froggatt-Nielsen parameter $\lambda_H$, which
arises from the breaking of the $U(1)_H$ 
symmetry and accounts for the hierarchies in the charged lepton and quark sectors 
of the Standard Model, arises from the vev of the $S_H$ field. It sets the scale here 
for the level of breaking of the $L_e-L_\mu-L_\tau$ symmetry. 

The collection of heavy neutrino fields is economical, with one pair of heavy 
lepton fields $N$ and $\bar{N}$ for each flavor generation. 
The lepton number violating scalar fields $S_{HL}$ and 
$\bar{S}_{HL}$ carry the horizontal symmetry charge. Upon 
receiving vacuum expectation values (vevs), they introduce a
scale  $M_{HL}=\langle S_{HL}\rangle \sim \langle \bar{S}_{HL}\rangle$ which then
via the seesaw mechanism sets the scale for $m$ in Eq.~(\ref{unperturb}).
Even after the spontaneous breaking of the $U(1)_L$ symmetry, the 
(accidental) $L_e-L_\mu-L_\tau$ symmetry of the $S_{HL}$ and 
$\bar{S}_{HL}$ fields interactions prevent contributions to the zero
entries in Eq.~(\ref{unperturb}).
A third scale is associated with vevs of the fields $S_H$ and $S_H^\prime$
and is small, ${\cal O}(\lambda _H)$, compared to $M_F$. 
It is this last scale which accounts for the 
$L_e-L_\mu-L_\tau$ violation. The $S_H$ and $S_H^\prime$ fields are singlets 
under $U(1)_L$, so their vevs cannot by themselves contribute to the  
zero entries in Eq.~(\ref{unperturb}). The $S_H$ and $S_H^\prime$ fields, the first of which 
is envisioned to play a role in the Froggatt-Nielsen mechanism in the quark 
sector, are assigned zero charges under $L_e-L_\mu-L_\tau$. 

\begin{table}[t]
\begin{tabular}{|c|r|r|r|r|r|}
\hline
       & $L_e$ & $L_\mu + L_\tau$ & $L_e-L_\mu -L_\tau$ & $L$ & $H$ \\ 
\hline
$L_{+1}$,$N_{+1}^{(i)}$    &   $+1$   &        $0$         &      $+1$           & $+1$ & $+1$\\ 
$\bar{N}_{-1}^{(i)}$ & $-1$ &      $0$         &      $-1$           & $-1$ & $-1$\\
\hline
$L_0^k$,$N_{0}^{(j)}$     &   $0$   &        $+1$         &      $-1$           & $+1$ & $0$ \\ 
$\bar{N}_{0}^{(j)}$ & $0$ &       $-1$         &      $+1$           & $-1$ & $0$ \\
\hline
\end{tabular}
\caption{Lepton number assignments for the fermion fields. The indices $i,j$ indicate the possibility of multiple 
generations of these fields. For three generations of light neutrinos, one takes $k=1,2$. The formulas 
presented in this paper correspond to the case $i=1$ and $j=1,2$.}
\label{lepton}
\end{table}

First consider the situation where the horizontal symmetry 
breaking is temporarily turned off (i.e. $\lambda_H=0$).
With the charges of the fields chosen as in Eq.~(\ref{def}), the
following symmetric mass matrix results:
\begin{widetext}
\begin{eqnarray}
{\cal M}_1&=&
\left (\begin{array}{ccccccccc}
0 & 0 & 0 & 0 & \phi_u & 0 & 0 & 0 & 0\\ 
0 & 0 & 0 & 0 & 0 & 0 & \phi_u & 0 & \phi_u \\
0 & 0 & 0 & 0 & 0 & 0 & \phi_u & 0 & \phi_u\\
0 & 0 & 0 & 0 & M_F & S_{HL} & 0 & S_{HL} & 0\\
\phi_u & 0 & 0 & M_F & 0 & 0 & \bar{S}_{HL} & 0 &  \bar{S}_{HL} \\
0 & 0 & 0 & S_{HL} & 0 & 0 & M_F & 0 & M_F \\
0 & \phi_u & \phi_u & 0 & \bar{S}_{HL} & M_F & 0 & M_F & 0 \\  
0 & 0 & 0 & S_{HL} & 0 & 0 & M_F & 0 & M_F \\
0 & \phi_u & \phi_u & 0 & \bar{S}_{HL} & M_F & 0 & M_F & 0 
\end{array}\right )\;.
\label{modelu}
\end{eqnarray}
\end{widetext}
The rows and columns are ordered as given in Eq.~(\ref{def}).
The entries $S_{HL}$ and $\bar{S}_{HL}$ represent the vevs of the fields.
As in all types of  Froggatt-Nielsen models there is understood 
to be an undetermined order one coefficient in front of every 
nonvanishing entry in this matrix since the symmetries only enforce 
that certain elements are zero but does by itself enforce any
relationships between nonzero elements apart from an overall scale. 
Some of the entries are 
related because the matrix is symmetric, but other entries which appear 
identical are only equivalent up to these coefficients (for example, $M_F$'s 
in the $4-5$ entry and in the $7-8$ entry).
We imagine the scale $M_{HL}$ is comparable to the scale $M_F$ (so that 
$\lambda_L\sim \lambda _H)$, whereas $M_H=\langle S_{H}\rangle \sim \langle S_{H}\prime\rangle$
is smaller (and ignored here temporarily).
The effect of assigning a nonzero $U(1)_H$ charge to the fields giving
rise to lepton number
violation, $S_{HL}$ and $\bar{S}_{HL}$, 
is to move the contributions off the diagonal, and
as a consequence generate the pseudo-Dirac structure in the effective 
light neutrino mass matrix.

The mass matrix in Eq.~(\ref{modelu}) gives rise to the 
light neutrino mass matrix $U_\nu$ in Eq.~(\ref{unperturb}) upon
integrating out the heavy sector. This mass matrix
has one maximal angle, one angle $\theta$
which is given in terms of (undetermined) 
order one coefficients in the full mass matrix in Eq.~(\ref{modelu}). 
Because the source of the light neutrino
masses arises from a limited number of interactions, we can define 
$\tan \theta_1$ to be the ratio of the $3-7$ and the $2-7$ 
entries in the matrix in Eq.~(\ref{modelu}). Likewise define
$\tan \theta_2$ to be the ratio of the $3-9$ and the $2-9$ 
entries. These two contributions add with in general differing 
weights, $m_1$ and $m_2$, to give the overall mixing angle $\theta$:
\begin{eqnarray}
m\sin \theta &=& m_1\sin \theta _1+m_2\sin \theta_2
\nonumber \\
m\cos \theta &=& m_1\cos \theta _1+m_2\cos \theta_2
\label{add}
\end{eqnarray}
The mass parameters $m_1$ and $m_2$ have complicated but completely 
known expressions in terms of the nonzero entries in Eq.~(\ref{modelu}).
If the second pair of superheavy neutrino fields $ N_0^{(2)}, \bar{N}_0^{(2)}$ is 
absent as in the model in Ref.~\cite{Berger:2006zw}, then there is necessarily
an alignment in angles $\theta =\theta_1$ (there is no angle $\theta_2$ in that 
case) which eventually extends to the 
the perturbative contributions to the light neutrino mass matrix (see Eqs.~(\ref{Goh})
and (\ref{entries}) below).
The overall scale $m$ is generated as a seesaw-type mass, and the 
undetermined ratio of order one coefficients has been expressed as the 
angle $\theta$ in Eq.~(\ref{add}). These terms break the lepton number 
symmetry but do not violate 
the $L_e-L_\mu -L_\tau$ symmetry.

Now we consider the small perturbations coming from the vevs of the fields 
$S_H$ snd $S_H^\prime$ whose interactions violate $L_e-L_\mu -L_\tau$ and
whose vevs at the same 
time as breaking the horizontal symmetry. Since these the Froggatt-Nielsen 
parameters are assumed to be small to explain the hierarchies in the fermion 
masses and mixings, they simultaneously account for the smallness of the 
perturbations to Eq.~(\ref{unperturb}). In fact, with these perturbations the 
experimental data, which requires a solar angles not maximal and a mass 
splitting between the two nonzero light neutrino masses (for the solar 
neutrino $\Delta m^2$), can be accomodated 
as a natural consequence of the Froggatt-Nielsen breaking of the horizontal 
symmetry.

\begin{table}[t]
\begin{tabular}{|c|r|r|r|r|r|}
\hline
       & $L_e$ & $L_\mu + L_\tau$ & $L_e-L_\mu -L_\tau$ & $L$ & $H$\\ 
\hline
$S_{HL}$    &   $-1$   &        $-1$         &      $0$       & $-2$ & $-1$\\ 
$\bar{S}_{HL}$    &   $+1$   &        $+1$         &      $0$ & $+2$ & $+1$\\ 
\hline 
$S_H$    &      $0$   &        $0$         &      $0$      & $0$ & $-1$ \\ 
$S_H^\prime$ &  $0$   &        $0$         &      $0$      & $0$ & $+1$ \\ 
\hline
\end{tabular}
\caption{Lepton numbers assigned to the scalar fields.}
\label{scalar}
\end{table}

First the effects of the field $S_H$ arise from the coupling 
$S_HN_{+1}\bar{N}_0^{(j)}$ which conserves the horizontal charge $U(1)_H$.
Upon spontaneous symmetry breaking the scalar field receives a vev and
gives contributions in elements of the mass matrix that were previously 
zero, 
\begin{widetext}
\begin{eqnarray}
{\cal M}_2&=&
\left (\begin{array}{ccccccccc}
0 & 0 & 0 & 0 & \phi_u & 0 & 0 & 0 & 0\\ 
0 & 0 & 0 & 0 & 0 & 0 & \phi_u & 0 & \phi_u \\
0 & 0 & 0 & 0 & 0 & 0 & \phi_u & 0 & \phi_u\\
0 & 0 & 0 & 0 & M_F & S_{HL} & PM_F & S_{HL} & PM_F\\
\phi_u & 0 & 0 & M_F & 0 & 0 & \bar{S}_{HL} & 0 &  \bar{S}_{HL} \\
0 & 0 & 0 & S_{HL} & 0 & 0 & M_F & 0 & M_F \\
0 & \phi_u & \phi_u & PM_F & \bar{S}_{HL} & M_F & 0 & M_F & 0 \\  
0 & 0 & 0 & S_{HL} & 0 & 0 & M_F & 0 & M_F \\
0 & \phi_u & \phi_u & PM_F & \bar{S}_{HL} & M_F & 0 & M_F & 0 
\end{array}\right )\;,
\label{model1}
\end{eqnarray}
\end{widetext}
where $P={\cal O}(M_H/M_F)={\cal O}(\lambda_H)$ is a small dimensionless 
parameter. Again apart from the matrix is symmetric, but apart from that 
relationship, the different instances of $P$ are related to each other by 
an undetermined order one coefficient. The $S_HN_{+1}\bar{N}_0^{(j)}$
interaction introduces a perturbation in the $L_e-L_\mu -L_\tau=2$ element
of Eq.~(\ref{unperturb}), so that it has the form,
\begin{eqnarray}
m^{}_{\ell 2}&=&m\left (\begin{array}{ccc}
z & \sin \theta & \cos \theta \\
\sin \theta & 0 & 0 \\
\cos \theta & 0 & 0
\end{array}\right )\;,
\label{perturb}
\end{eqnarray}
where the small quantity $z$ is of order the perturbation,
$P$. So whatever the number of fields $\bar{N}_0^{(j)}$ indexed by
$j$ these will contribute only to the $1-1$ entry of the light
neutrino mass matrix. These contributions must necessarily involve
the lepton-number breaking vev of the $S_{HL}$ field as well as a 
Froggatt-Nielsen suppression of order $P$.

In addition to this perturbation, there is a field $S_H^\prime$ 
with $U(1)_H$ charge $+1$, then it provides a coupling
$S_H^\prime N_{0}^{(j)}\bar{N}_{-1}$. Since this interaction has charge $L_e-L_\mu -L_\tau=-2$, it will 
produce perturbations in the
$2-3$ subblock in the light neutrino mass matrix. After $S_H^\prime$
receives a vev, it introduces contributions to certain entries in 
Eq.~(\ref{model1}) that were previously zero,
\begin{widetext}
\begin{eqnarray}
{\cal M}_3&=&
\left (\begin{array}{ccccccccc}
0 & 0 & 0 & 0 & \phi_u & 0 & 0 & 0 & 0\\ 
0 & 0 & 0 & 0 & 0 & 0 & \phi_u & 0 & \phi_u \\
0 & 0 & 0 & 0 & 0 & 0 & \phi_u & 0 & \phi_u\\
0 & 0 & 0 & 0 & M_F & S_{HL} & PM_F & S_{HL} & PM_F\\
\phi_u & 0 & 0 & M_F & 0 & P^\prime M_F & \bar{S}_{HL} & P^\prime M_F &  \bar{S}_{HL} \\
0 & 0 & 0 & S_{HL} & P^\prime M_F & 0 & M_F & 0 & M_F \\
0 & \phi_u & \phi_u & PM_F & \bar{S}_{HL} & M_F & 0 & M_F & 0 \\  
0 & 0 & 0 & S_{HL} & P^\prime M_F & 0 & M_F & 0 & M_F \\
0 & \phi_u & \phi_u & PM_F & \bar{S}_{HL} & M_F & 0 & M_F & 0 
\end{array}\right )\;.
\label{model3}
\end{eqnarray}
\end{widetext}
Here the perturbation $P^\prime={\cal O}(\lambda_H)$ is naturally small
by the Froggatt-Nielsen mechanism.  The effective light neutrino mass matrix in 
Eq.~(\ref{unperturb}) becomes,
\begin{eqnarray}
m^{}_{\ell 3}&=&m\left (\begin{array}{ccc}
z & \sin \theta & \cos \theta \\
\sin \theta & y & d \\
\cos \theta & d & x 
\end{array}\right )\;,
\label{Goh}
\end{eqnarray}
where the small quantities $x,y,d$ are of order the perturbation,
$P^\prime$. The relationship between the four terms in the $2-3$ subblock
of this matrix is enforced by the restricted nature of the couplings of the 
lepton doublets $L$ to the heavy neutrino fields $N$ and $\bar{N}$.
Furthermore the parameters are related to each other so that the $3\times 3$ matrix 
$m_{\ell 3}$ has a vanishing determinant and therefore there is one massless 
neutrino. This condition is enforced by the symmetries of the model. The form in Eq.~(\ref{Goh}) yields
a light neutrino mass matrix that can be phenomenologically acceptable for
describing all the data from atmospheric and solar neutrino oscillations.
In fact the quantities $x,y,d$ have the form
\begin{eqnarray}
x&=&z^\prime_{11}\cos^2 \theta_1+z^\prime_{22}\cos^2 \theta_2+z^\prime_{12}\cos \theta_1\cos \theta_2 
\nonumber \\
y&=&z^\prime_{11}\sin^2 \theta_1+z^\prime_{22}\sin^2 \theta_2+z^\prime_{12}\sin \theta_1\sin \theta_2 
\nonumber \\
d&=&z^\prime _{11}\sin\theta_1\cos \theta_1+z^\prime_{22}\sin\theta_2\cos \theta_2
\nonumber \\
&&+\half z^\prime _{12}(\sin\theta_1\cos \theta_2+\sin\theta_2\cos \theta_1)\;.
\label{entries}
\end{eqnarray}
The quantities $z^\prime_{ij}$ are of order $P^\prime$ in the perturbation. The formulas in Eq.(\ref{entries})
have an obvious generallization to the case where more (than two) superheavy fields  ($N_0^{(j)}, \bar{N}_0^{(j)}$,
$j=1,2,3,\dots$) are present. On the other hand, when only one such superheavy field is present,  
$\theta_1 \equiv \theta$ and the terms involving $\theta _2$ are absent. Then the perturbations can be seen 
to align necessarily with the atmospheric mixing angle $\theta$. This yields an unacceptable value of $U_{e3}=0$
(see Eqn.~(\ref{quant}) below). 

The more constrained model presented in Ref.~\cite{Berger:2006zw} has sometimes been categorized as an 
example of the Simple Real Scaling (SRS) ansatz\cite{Lavoura:2000kg,Mohapatra:2006xy}. However, that 
constrained model falls into the SRS class of theories by accident (i.e. the SRS form is not required by the 
symmetries). As a consequence of being an SRS model,
it indeed exhibits an unacceptable vanishing third neutrino mixing angle. The 
generalization to the model described in this paper eliminates this phenomenological prediction (while keeping
a massless neutrino), and puts the new model outside the SRS 
class\cite{Blum:2007qm,Joshipura:2009fu,Goswami:2009bd,Hettmansperger:2011bt,Adhikary:2012kb,Chakraborty:2014hfa,Ghosal:2015lwa,Samanta:2015hxa,
Samanta:2016wca}.

\section{Physical predictions of the model}

This mass matrix in Eq.~(\ref{Goh}), involving perturbations $x,y,z,d$ collectively of order $\delta$, 
can be related to the physical observables\cite{Goh:2002nk}. The atmospheric neutrino oscillation 
parameters are
\begin{eqnarray}
\sin^2 2\theta_{A}&=&\sin^2 2\theta + {\cal O}(\delta^2)\;, 
\nonumber \\
\triangle m_A^2&=&-m^2 + 2 \triangle m^2_{\odot}+ {\cal O}(\delta^2)\;.
\label{first}
\end{eqnarray}
The solar neutrino parameters are given by
\begin{eqnarray}
\sin^22\theta_{\odot}&=&1-\left (\frac{\triangle m_\odot^2}{4\triangle
m_A^2}-z\right )^2+{\cal O}\;, 
\nonumber \\
R&=& \frac{\triangle m_\odot^2}{\triangle
m_A^2}=2(z+\vec{v}\cdot\vec{x})+{\cal O}(\delta^2)\;.
\label{second}
\end{eqnarray} 
These formulas give the leading perturbations away from leading order ones in
Eq.~(\ref{unperturb_angles}).
Finally one has
\begin{eqnarray}
U_{e3}&=&\vec{A}\cdot(\vec{v}\times\vec{x})+{\cal O}(\delta^3)\;.
\end{eqnarray}
where $\vec{v}=(\cos^2\theta,\sin^2\theta,\sqrt{2}\sin\theta\cos\theta)$, 
$\vec{x}=(x,y,\sqrt{2}d)$ and
$\vec{A}~=~\frac{1}{\sqrt{2}}(1,1,0)$. 
Full unapproximated expressions can be obtained by using the results of Ref.~\cite{Adhikary:2013bma}.

From Eq.~(\ref{entries}) we have
\begin{eqnarray}
\vec{v}\cdot\vec{x}&=&z^\prime_{11}\cos^2 \alpha _1+z^\prime_{22}\cos^2 \alpha _2
+z^\prime_{12}\cos \alpha_1 \cos \alpha_2
\nonumber \\
U_{e3}&=&z^\prime_{11}\sin \alpha_1 \cos \alpha _1+z^\prime_{22}\sin \alpha _2\cos \alpha _2
\nonumber \\
&&+\half z^\prime_{12}\sin (\alpha_1 +\alpha_2)\;,
\label{quant}
\end{eqnarray}
where $\alpha _i=\theta - \theta _i$ are the angles measuring the nonalignment of the 
contributions from each superheavy neutrino pair with the total mixing angle $\theta$
between the second and third generations in the light neutrino mass matrix (Eq.~(\ref{Goh})).
It is straightforward to find parameter values which produce observables in acceptable 
agreement with experiment.
Again there is an obvious generalization to Eqs.~(\ref{add}) and (\ref{quant}) when extra superheavy 
neutrinos are included, but the overall mass pattern remains the same.

The light neutrino mass matrix describes one massless neutrino, so it automatically satisfies the condition 
$\det M_\nu = 0$\cite{Minkowski:1977sc,Branco:2002ie}. This condition has been exploited in Frogatt-Nielsen models previously, as an supplementary constraint 
on the parameters\cite{Kaneta:2017byo} of a model with a normal mass hierarchy.

Dirac phases can be easily included in the model. However the symmetries do not constrain
these phases, so no unique prediction for CP violation in neutrino oscillations can be made, beyond 
the expected overall scale associated with the product of the three nonzero mixing angles.

\section{Summary and Conclusions}

We have presented a model where lepton number violation arises from the vev of a scalar field which also 
carries a horizontal symmetry charge. The large mixing angles observed in neutrino oscillations experiments
arise from an $L_e-L_\mu -L_\tau$ symmetry together with Froggatt-Nielsen neutrino anarchy in the second and third 
generations. Froggatt-Nielsen symmetry breaking then accounts for the smaller observables.
The resulting light neutrino physics has an inverted hierarchy, two 
large and one small mixing angle, and one exactly massless neutrino. There is 
a nonzero expectation for neutrinoless double beta decay (the mass parameter measured
in those experiments is $mz$) as is familiar in inverted hierarchy models.
The form of the light neutrino mass matrix, determined at a high energy scale where 
the spontaneous symmetry breaking occurs,
and its phenomenology are unaffected under renormalization group evolution. Increasing the number 
of heavy neutrino fields beyond the minimum number of three generations yields the same overall
phenomenology. However, the case of three generations with two having exactly the same 
charge assignments is sufficient to provide consistency with the experimental neutrino data. 

One can take the point of view that a full three-generational anarchy in Froggatt-Nielsen charges 
is compatible with the neutrino data, and the small quantities are just accounted for by small parameters
and/or cancellations. This is certainly reasonable since the hierarchy in the neutrino observables is 
quite mild. On the other hand, the model presented here offers a reason for their smallness in parallel
with the quark and charge lepton hierarchies, namely the same Froggatt-Nielsen mechanism.

\end{document}